\UseRawInputEncoding
\documentclass[11pt]{article}

\usepackage{cite}
\usepackage{tikz}
\usetikzlibrary{arrows.meta}
\usetikzlibrary{positioning}

\usepackage{amsmath,amssymb}
\usepackage{amsthm}
\usepackage{graphicx}
\usepackage{algorithm}
\usepackage{algorithmic}
\usepackage{listings}
\usepackage{xcolor}
\usepackage[colorlinks=true,linkcolor=blue,citecolor=blue,urlcolor=blue]{hyperref}
\usepackage{enumitem}
\usepackage[T1]{fontenc}
\usepackage[nameinlink,capitalise]{cleveref}

\lstdefinestyle{hqpefcode}{%
  language=Python,
  basicstyle=\ttfamily\small,
  numbers=left,
  numberstyle=\tiny,
  stepnumber=1,
  numbersep=5pt,
  showstringspaces=false,
  tabsize=4,
  breaklines=true,
  breakatwhitespace=true,
  keywordstyle=\color{blue}\bfseries,
  commentstyle=\color{gray}\itshape,
  stringstyle=\color{red}
}
\begin{document}

\title{HQPEF-Py: Metrics, Python Patterns, and Guidance for Evaluating Hybrid Quantum Programs}
\author{\ Michael Adjei Osei\\ Sidney Shapiro\\[1ex]
{\normalsize University of Lethbridge, 4401 University Dr W, Lethbridge, AB T1K 6T4, Canada}}
\date{ }
\maketitle

\begin{abstract}
We study how to evaluate hybrid quantum programs as end-to-end workflows rather than as isolated devices or algorithms. Building on the Hybrid Quantum Program Evaluation Framework (HQPEF), we formalize a workflow-aware Quantum Readiness Level (QRL) score; define a normalized speedup under quality constraints for the Utility of Quantumness (UQ); and provide a timing-and-drift audit for hybrid pipelines. We complement these definitions with concise Python reference implementations that illustrate how to instantiate the metrics and audit procedures with state-of-the-art classical and quantum solvers (e.g., via Qiskit~\cite{kanazawa2023qiskit} or PennyLane~\cite{Bergholm2018}), while preserving matched-budget discipline and reproducibility.
\end{abstract}

\section{Introduction}
Hybrid quantum--classical workflows in the NISQ era~\cite{Preskill2018} demand more than peak-device metrics (e.g., IBM's quantum volume~\cite{Cross2019}) as measures of progress. Teams must track end-to-end maturity, benchmark against strong classical baselines under matched resources, and surface non-obvious bottlenecks (e.g., encoding or error mitigation) that dominate wall-clock time. 

At a high level, our perspective is that a hybrid quantum program should be
evaluated as an \emph{end-to-end workflow} rather than as an isolated
device or algorithm. The central idea is to separate questions of
\emph{maturity} (how well-governed and reproducible the workflow is) from
questions of \emph{utility} (whether a given hybrid stack delivers an
advantage under matched resources) and from \emph{engineering bottlenecks}
(where time and instability concentrate along the pipeline). This separation
is important for at least three reasons: it makes reported ``quantum
advantage'' claims more comparable across studies, it provides concrete
targets for improving real systems (e.g., addressing encoding or
transpilation rather than only hardware), and it gives non-technical
stakeholders a structured way to reason about readiness and risk. The
QRL, UQ, and workflow-audit constructs we introduce are designed to support this layered view of evaluation.

This paper develops the underlying evaluation constructs of HQPEF and shows how they can be realized in Python as lightweight, auditable components. Rather than proposing a monolithic toolkit, we focus on theory, reference code samples, and practical direction for implementing hybrid-quantum evaluation workflows.

Existing benchmarking efforts focus primarily on device-level figures of merit
or algorithm-specific performance curves; they do not directly address workflow
maturity, governance, and fair comparison under matched resources.

The remainder of this paper is organized as follows.
\Cref{sec:readiness} formalizes the QRL scoring rule,
\Cref{sec:uq} introduces the UQ metric and protocol,
\Cref{sec:methodology} summarizes methodological guidance,
\Cref{sec:audit} describes workflow auditing,
\Cref{sec:experiments} presents a synthetic case study and Python patterns,
and the appendix collects reference code.

\textbf{Contributions.}
(1) A formal QRL rubric mapping evidence-backed checklists to staged readiness levels (1--9)~\cite{NASA2017}. 
(2) A solver-agnostic UQ definition that decouples solution quality from runtime via a normalized speedup at a target quality. 
(3) A workflow audit that attributes latency and instability across stages and returns a Pareto-style bottleneck set. 
(4) Python reference implementations and usage patterns that demonstrate how these constructs can be integrated into real hybrid workflows.

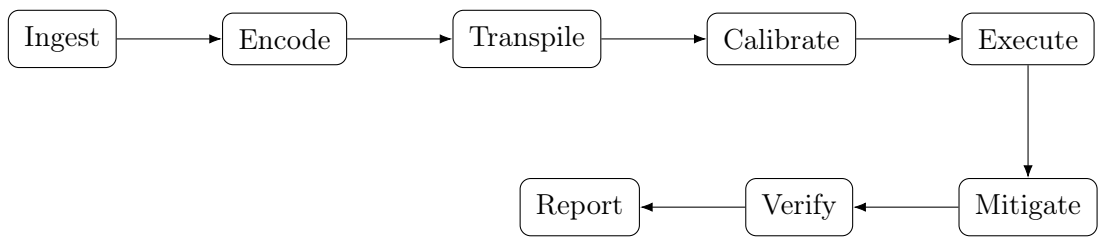
\begin{figure}[htbp]
\centering
\begin{tikzpicture}[node distance=1.5cm and 1.4cm, every node/.style={rounded corners, draw, align=center, inner sep=6pt}]
\node (ingest) {Ingest};
\node (encode) [right=of ingest] {Encode};
\node (transpile) [right=of encode] {Transpile};
\node (calib) [right=of transpile] {Calibrate};
\node (exec) [right=of calib] {Execute};
\node (mitigate) [below=of exec] {Mitigate};
\node (verify) [left=of mitigate] {Verify};
\node (report) [left=of verify] {Report};

\draw[-{Latex}] (ingest) -- (encode);
\draw[-{Latex}] (encode) -- (transpile);
\draw[-{Latex}] (transpile) -- (calib);
\draw[-{Latex}] (calib) -- (exec);
\draw[-{Latex}] (exec) -- (mitigate);
\draw[-{Latex}] (mitigate) -- (verify);
\draw[-{Latex}] (verify) -- (report);

\node[draw=none, align=left, above=0.9cm of encode, inner sep=0pt] (labels) {
\textbf{Evaluation modules:}\\
\quad \texttt{qrl}: readiness scoring\\
\quad \texttt{uq}: matched-budget benchmarking\\
\quad \texttt{workflow}: stage timing \& drift\\
\quad \texttt{stakeholders}: reporting views
};
\end{tikzpicture}
\caption{Conceptual instrumentation of a hybrid pipeline.\\
Hooks capture per-stage timings and calibration drift, while the benchmarking harness enforces resource matching and strong classical baselines.}
\label{fig:pipeline}
\end{figure}

\section{Workflow Readiness: Formalization and Scoring}
\label{sec:readiness}
Let a checklist vector be $c \in \{0,1\}^k$ with components for problem formulation, encoding specification, prototype components, integrated pipeline, reproducible runs, error budget, classical baseline, matched-budget protocol, change control, audit trail, service-level objectives (SLOs), external replication, and governed deployment. Let $\delta \in \mathbb{R}_{\ge 0}$ denote long-term calibration drift (parts-per-million, ppm). We define a transparent readiness score:
\begin{equation}
S(c,\delta) = \sum_{i=1}^{k} w_i c_i \;+\; g(\delta), 
\qquad 
g(\delta) = 
\begin{cases}
10, & 0 < \delta \le 10,\\
6, & 10 < \delta \le 100,\\
0, & \text{otherwise.}
\end{cases}
\label{eq:score}
\end{equation}
The QRL stage is then obtained by a piecewise map:
\begin{equation}
\text{QRL}(S) = 
\begin{cases}
1, & S < 10;\\
2, & 10 \le S < 20;\\
3, & 20 \le S < 35;\\
4, & 35 \le S < 50;\\
5, & 50 \le S < 65;\\
6, & 65 \le S < 75;\\
7, & 75 \le S < 85;\\
8, & 85 \le S < 95;\\
9, & S \ge 95~,
\end{cases}
\label{eq:qrl}
\end{equation}
which yields an integer readiness level $1 \le \text{QRL}(S) \le 9$.

In practice, the weights $w_i$ in \eqref{eq:score} are configured to reflect the governance and maturity priorities of a project or organization. A regulated deployment may assign higher weights to governed deployment, external replication, and audit trail, whereas an early-stage research prototype may prioritize encoding specification and prototype components. The rubric and weights should be versioned so that changes in QRL can be interpreted in light of the underlying definitions.

\paragraph{Example (QRL scoring).}
Consider a simplified rubric with $k=5$ checklist items:
problem formulation, encoding specification, integrated pipeline, classical baseline, and audit trail. Suppose the weights are
\[
w = (8, 8, 10, 6, 6)
\]
and a given project has satisfied
problem formulation, encoding specification, and classical baseline but not yet implemented an integrated pipeline or audit trail:
\[
c = (1, 1, 0, 1, 0).
\]
Let the long-term calibration drift be $\delta = 12$~ppm. Then $g(\delta)=6$ by \eqref{eq:score}, and
\[
S(c,\delta) = 8\cdot 1 + 8\cdot 1 + 10\cdot 0 + 6\cdot 1 + 6\cdot 0 + 6
= 28.
\]
From \eqref{eq:qrl}, $20 \le S < 35$ implies $\text{QRL}(S)=3$.
If the team later implements a minimal integrated pipeline (updating $c_3$ from $0$ to $1$) and reduces drift to $\delta = 8$~ppm, the score becomes
$S=38$ and the readiness level advances to $\text{QRL}(S)=4$.

\section{Utility of Quantumness (UQ) Under Matched Budgets}
\label{sec:uq}
Let a resource budget be $B = (t, c, e)$ for time ($s$), cost (USD), and energy (J). Let $Q(\theta) \in [0,1]$ (higher is better) be a quality metric on solver output $\theta$. Given two solvers \textsf{A} and \textsf{B} evaluated on the same instance family under identical budgets, we report the \emph{normalized speedup at target quality $\tau$} as 
\begin{equation}
S_{\text{norm}}(\tau) = \frac{\min\{T_A : Q(\theta_A) \ge \tau\}}{\min\{T_B : Q(\theta_B) \ge \tau\}}~,
\label{eq:snorm}
\end{equation}
where $T_{\cdot}$ denotes the wall-clock runtime. This decouples speed from absolute quality and discourages cherry-picking of single runs in comparisons~\cite{Ronnow2014}. 

\noindent \textbf{Reference protocol.} 
\begin{enumerate}[label=(\arabic*), leftmargin=*]
  \item Choose an instance generator and quality metric $Q(\cdot)$.
  \item Preregister budgets $B$ and target $\tau$.
  \item Run the named solvers with fixed random seeds.
  \item Summarize performance by reporting $E[Q]$, $E[T]$, $E[\text{energy}]$, $E[\text{cost}]$ (means over instances), and provide full outcome distributions.
  \item Report $S_{\text{norm}}(\tau)$ with confidence intervals, where applicable.
\end{enumerate}

\paragraph{Example (computing $S_{\text{norm}}$).}
Suppose solvers \textsf{A} and \textsf{B} are evaluated on the same three instances under an identical budget $B$. For each instance we record the best quality achieved under the budget and the corresponding runtime (in milliseconds):

\begin{center}
\begin{tabular}{c|ccc|ccc}
\hline
 & \multicolumn{3}{c|}{Solver \textsf{A}} & \multicolumn{3}{c}{Solver \textsf{B}} \\
Instance & $T_A$ & $Q_A$ &  & $T_B$ & $Q_B$ & \\
\hline
1 & 12 & 0.72 & &  9 & 0.70 \\
2 & 15 & 0.75 & & 10 & 0.73 \\
3 & 11 & 0.68 & &  8 & 0.69 \\
\hline
\end{tabular}
\end{center}

Take a target quality $\tau = 0.72$. For solver \textsf{A}, instances 1 and 2 meet the threshold, with runtimes $12$ and $15$, so
$\min\{T_A : Q_A \ge \tau\} = 12$.
For solver \textsf{B}, only instance 2 meets the threshold, with runtime $10$, so
$\min\{T_B : Q_B \ge \tau\} = 10$.
Thus
\[
S_{\text{norm}}(0.72) = \frac{12}{10} = 1.2.
\]
In this convention, $S_{\text{norm}}(\tau) > 1$ indicates that solver \textsf{B} reaches quality $\tau$ faster than solver \textsf{A} (smaller denominator), while $S_{\text{norm}}(\tau) < 1$ would indicate the opposite. If either solver never reaches quality $\tau$ under the budget, the corresponding set in \eqref{eq:snorm} is empty and $S_{\text{norm}}(\tau)$ is treated as undefined or $+\infty$.

\section{Methodological Context and Guidance}
\label{sec:methodology}
\paragraph{Choosing $Q(\cdot)$ and $\tau$.}
Select a monotonic, bounded quality metric $Q \in [0,1]$ that is calibrated to the task and (if feasible) invariant to instance size (e.g., a normalized cut value for MaxCut). Predeclare the target threshold $\tau$ to prevent post-hoc threshold shopping. 

\paragraph{Budgets and fairness.}
Fix the budget tuple $B = (t,c,e)$ jointly; if only time is strictly enforced in practice, report the energy and cost realized anyway and state any assumptions. Always ensure strong classical baselines and release solver configurations for reproducibility. 

\paragraph{Instance design.}
Prefer an instance generator with adjustable difficulty parameters (e.g., size, sparsity, noise). Stratify instances by difficulty to stabilize estimates, and disclose the stratification scheme used. 

\paragraph{Uncertainty and reporting.}
Report distributional summaries of outcomes (not just means), include bootstrap confidence intervals for $S_{\text{norm}}(\tau)$, and examine sensitivity to $\tau$ through a performance sweep (see Listing~\ref{lst:tau_sweep}). 

\paragraph{Audit usage.}
Use per-stage time shares to prioritize engineering effort. For example, if encoding and transpilation stages dominate overall latency, hardware improvements alone may not significantly reduce end-to-end runtime.

\paragraph{Governance.}
Record random seeds, software versions, and change logs. Publish an audit trail artifact and a preregistration note. These steps align with the definitions and protocol formalized in Eqs.~(\ref{eq:score})--(\ref{eq:snorm}) and Listing~\ref{lst:audit_fn}.

\begin{lstlisting}[style=hqpefcode,caption={Benchmark harness API (abridged).},label={lst:tau_sweep}]
from dataclasses import dataclass
from typing import Callable, Any, Dict, List, Optional

@dataclass
class Budget:
    time_s: float
    cost_usd: float = 0.0
    energy_j: float = 0.0

@dataclass
class Result:
    solver: str
    instance_id: int
    quality: float
    time_s: float
    energy_j: float
    cost_usd: float
    meta: Dict[str, Any]

class BenchmarkRunner:
    def __init__(self, solvers: Dict[str, Callable],
                 instance_gen: Callable[[int], Any],
                 target_quality: Optional[float] = None):
        self.solvers = solvers
        self.instance_gen = instance_gen
        self.target_quality = target_quality

    def run(self, num_instances: int, budget: Budget, **kwargs) -> List[Result]:
        ...
\end{lstlisting}

\section{Workflow Audit and Bottleneck Attribution}
\label{sec:audit}
For each run, we observe a vector of stage durations $d \in \mathbb{R}_{\ge 0}^m$ across $m$ stages (ingest~$\to$~report). With $R$ replicates $D = \{d^{(r)}\}_{r=1}^R$, define the mean share per stage $i$ as 
\[
\bar{s}_i = \frac{1}{R} \sum_{r=1}^{R} \frac{d^{(r)}_i}{\sum_{j=1}^{m} d^{(r)}_j}~.
\]
We return the top-$k$ stages ranked by $\bar{s}_i$ as the bottlenecks, and we summarize the calibration drift by reporting the mean and empirical 95th-percentile of the observed drift samples.

\paragraph{Example (stage-share audit).}
Assume $m=3$ stages labeled encode, transpile, and execute. Two runs produce the following durations (in milliseconds):
\[
d^{(1)} = (20, 50, 30), \qquad
d^{(2)} = (15, 45, 40).
\]
For run~1, the total time is $100$, so the shares are $(0.20, 0.50, 0.30)$. For run~2, the total is also $100$, so the shares are $(0.15, 0.45, 0.40)$. Averaging the shares across runs yields
\[
\bar{s}_{\text{encode}} = 0.175,\quad
\bar{s}_{\text{transpile}} = 0.475,\quad
\bar{s}_{\text{execute}} = 0.35.
\]
If we report the top-$k$ stages with $k=2$, the bottleneck set is
$\{\text{transpile}, \text{execute}\}$; engineering effort should focus there rather than on encoding.

\section{Reference Python Patterns}
\label{sec:patterns}
This section sketches how practitioners can integrate the HQPEF constructs into their own code bases. The goal is not to prescribe a specific package layout, but to offer reusable patterns for hybrid workflow evaluation.

A common pattern is to implement a callable
\texttt{fn(instance, budget, seed) -> (quality, information)} that wraps a given solver (classical or quantum), enforces a time budget, and returns both a normalized quality score and metadata such as runtime, energy, or cost. The listings in the appendix show how to structure these components in a way that is composable and testable. In particular, the \texttt{BenchmarkRunner} in Listing~\ref{lst:bench_harness} centralizes instance generation, budgeting, and result collection, while utility functions such as \texttt{audit\_bottlenecks} in Listing~\ref{lst:audit_fn} implement the workflow audit.

\section{Experimental Demonstration (Synthetic)}
\label{sec:experiments}
We include a toy benchmark on a binary quadratic problem (with $n=24$ variables and a time budget of 50~ms per instance) that compares a classical simulated annealer with a greedy ``quantum-like'' heuristic. The aim is to validate the harness logic and reporting (means, 95th percentile, and $S_{\text{norm}}$ at the declared $\tau$) rather than to claim solver superiority. Full outcome distributions, random seeds, and solver configurations are provided in the accompanying code.

\noindent \textbf{Reporting $S_{\text{norm}}(\tau)$.} For a preregistered $\tau$ (e.g., $0.70$), report the minimum time each solver achieves $Q \ge \tau$ over the set of instances, then compute $S_{\text{norm}}$ via Eq.~(\ref{eq:snorm}). Include bootstrap confidence intervals if appropriate, for example using the paired-bootstrap procedure in Listing~\ref{lst:bootstrap_ci}.

\section{Reproducibility and Governance}
Our recommended practice favors determinism (user-set seeds), versioned configurations, and change control. We recommend preregistering $\tau$ and budgets $B$, releasing instance generators, and enabling third-party replication. The QRL rubric (definitions of readiness levels) should be fixed and documented in code for auditability.

\section{Limitations}
The Python examples use minimal solvers; in applied settings, teams should integrate state-of-the-art classical and quantum backends for decision-grade comparisons. Metric design assumes a monotonic quality measure in $[0,1]$; each application domain must appropriately calibrate $Q(\cdot)$ and verify cross-instance comparability. Real energy and cost accounting requires proper metering or provider usage exports.

\section{Related Work}
We build on standard treatments of quantum computing~\cite{Nielsen2010}, the NISQ perspective~\cite{Preskill2018}, variational algorithms for near-term devices~\cite{Cerezo2021}, quantum annealing paradigms~\cite{Albash2018}, and modern toolchains for quantum programming~\cite{kanazawa2023qiskit,Bergholm2018}. Various benchmarking suites for quantum devices and algorithms have been proposed, focusing on instance design and device-level robustness. Our contribution is orthogonal to these efforts: we provide a workflow- and governance-oriented evaluation layer, together with concrete Python patterns for implementing matched-budget benchmarking and workflow audits around such suites.

\section{Formal Properties and Proofs}
The following statements formalize properties of the scoring function, stage mapping, normalized speedup, and audit shares defined in the main text.

\noindent \textbf{Proposition 1 (Determinism and monotonicity of readiness).} Let $S(c,\delta)$ be defined as in Eq.~(\ref{eq:score}) with weights $w_i \ge 0$ and $c_i \in \{0,1\}$. Then $S$ is deterministic given $(c,\delta)$ and is nondecreasing in each $c_i$ and in $\delta$ within each drift bracket of $g(\delta)$. Consequently, the stage map $\text{QRL}(S)$ in Eq.~(\ref{eq:qrl}) is a nondecreasing step function of $S$.

\textit{Proof.} $S(c,\delta) = \sum_i w_i c_i + g(\delta)$ is a sum of a linear form in $c$ with nonnegative coefficients plus a piecewise-constant function $g(\delta)$ defined on disjoint $\delta$-intervals. Increasing any $c_i$ from 0 to 1 increases $S$ by $w_i \ge 0$. For a fixed drift bracket, $g(\delta)$ remains constant; moving to a higher drift bracket never decreases $g$. Thresholding $S$ to map to a QRL stage preserves monotonicity in $S$. \hfill$\square$

\medskip
\noindent \textbf{Proposition 2 (Unit invariance of $S_{\text{norm}}(\tau)$).} For any $\tau \in [0,1]$, the normalized speedup $S_{\text{norm}}(\tau) = \min\{T_A : Q(\theta_A) \ge \tau\} \,/\, \min\{T_B : Q(\theta_B) \ge \tau\}$ is invariant under any common positive rescaling of the time unit.

\textit{Proof.} Multiplying all times by a constant factor $\alpha > 0$ multiplies both the numerator and denominator of $S_{\text{norm}}(\tau)$ by $\alpha$, leaving the ratio unchanged. \hfill$\square$

\medskip
\noindent \textbf{Proposition 3 (Dominance implies $S_{\text{norm}}(\tau) \le 1$).} If solver $A$ stochastically dominates solver $B$ in time-to-$\tau$ (i.e., $T_A \le T_B$ almost surely across instances), then $S_{\text{norm}}(\tau) \le 1$.

\textit{Proof.} Under stochastic dominance, $\min\{T_A : Q(\theta_A) \ge \tau\} \le \min\{T_B : Q(\theta_B) \ge \tau\}$ for each instance, hence their ratio is at most 1. \hfill$\square$

\medskip
\noindent \textbf{Lemma 1 (Stage-share normalization).} With per-run stage shares $s^{(r)}_i = d^{(r)}_i / \sum_j d^{(r)}_j$, we have $\sum_i s^{(r)}_i = 1$ for each run $r$, and $\sum_i \bar{s}_i = 1$ for their averages $\bar{s}_i$.

\textit{Proof.} By construction, the $s^{(r)}_i$ for a given run sum to 1. Averaging these shares over $R$ runs yields $\bar{s}_i = \frac{1}{R}\sum_r s^{(r)}_i$, and summing $\bar{s}_i$ over $i$ gives $\sum_i \bar{s}_i = \frac{1}{R}\sum_r \sum_i s^{(r)}_i = \frac{1}{R}\sum_r 1 = 1$. \hfill$\square$

\section{Conclusion and Future Directions}
We have presented HQPEF-Py as a collection of evaluation metrics, formal properties, and Python reference patterns for assessing hybrid quantum programs. By formalizing a workflow-aware Quantum Readiness Level, enforcing matched-budget benchmarking via normalized speedup at target quality, and instrumenting per-stage timing and calibration drift, the framework shifts attention from device-centric figures of merit to application-grounded maturity, fairness, and reproducibility.

The Python examples illustrate one way to instantiate these ideas in practice without committing to a specific toolkit or stack. Future work includes richer domain-specific quality metrics, improved energy and cost metering, adapters for emerging quantum and classical backends, and standardized benchmark suites and audit artifacts suitable for regulated deployments and artifact-evaluation tracks.

\bibliographystyle{unsrt}
\bibliography{ref}

\appendix
\renewcommand{\thesection}{\Alph{section}}
\renewcommand{\thesubsection}{\Alph{section}.\arabic{subsection}}

\section{Reference Code Examples (Python)}
This appendix provides executable examples that implement the core metrics, the audit routine, and end-to-end benchmarking logic described in the main text.

\subsection{Core metrics: readiness, stage mapping, and \texorpdfstring{$S_{\text{norm}}(\tau)$}{Snorm(tau)}}
\begin{lstlisting}[language=Python, caption={Benchmark harness API (expanded).}, label={lst:bench_harness}]
from dataclasses import dataclass
from typing import Callable, Any, Dict, List, Optional, Tuple

@dataclass
class Budget:
    time_s: float
    cost_usd: float = 0.0
    energy_j: float = 0.0

@dataclass
class Result:
    solver: str
    instance_id: int
    quality: float
    time_s: float
    energy_j: float
    cost_usd: float
    meta: Dict[str, Any]

class BenchmarkRunner:
    def __init__(self,
                 solvers: Dict[str, Callable[[Any, Budget, int],
                                            Tuple[float, Dict[str, Any]]]],
                 instance_gen: Callable[[int, int], Any],
                 target_quality: Optional[float] = None):
        self.solvers = solvers
        self.instance_gen = instance_gen
        self.target_quality = target_quality

    def run(self, num_instances: int, budget: Budget, seed: int = 7) -> List[Result]:
        import numpy as np
        rng = np.random.default_rng(seed)
        out: List[Result] = []
        for i in range(num_instances):
            inst = self.instance_gen(i, int(rng.integers(1e9)))
            for name, fn in self.solvers.items():
                q, info = fn(inst, budget, int(rng.integers(1e9)))
                out.append(Result(
                    solver=name, instance_id=i, quality=float(q),
                    time_s=float(info.get("time_s", float("nan"))),
                    energy_j=float(info.get("energy_j", 0.0)),
                    cost_usd=float(info.get("cost_usd", 0.0)),
                    meta=info
                ))
        return out
\end{lstlisting}

\begin{lstlisting}[language=Python, caption={Core metrics implementing Eqs.~(1)--(3).}, label={lst:core_metrics}]
from __future__ import annotations
from typing import Iterable, Tuple, List
import numpy as np, math

def qrl_score(c: Iterable[int], w: Iterable[float], delta_ppm: float) -> float:
    c = np.asarray(list(c), dtype=float)
    w = np.asarray(list(w), dtype=float)
    s = float(np.dot(w, c))
    if 0 < delta_ppm <= 10:
        g = 10.0
    elif 10 < delta_ppm <= 100:
        g = 6.0
    else:
        g = 0.0
    return s + g

def qrl_stage(S: float) -> int:
    if S < 10: return 1
    if S < 20: return 2
    if S < 35: return 3
    if S < 50: return 4
    if S < 65: return 5
    if S < 75: return 6
    if S < 85: return 7
    if S < 95: return 8
    return 9

def normalized_speedup_at_tau(
    times_A: Iterable[float],
    quals_A: Iterable[float],
    times_B: Iterable[float],
    quals_B: Iterable[float],
    tau: float
) -> float:
    tA = [t for t, q in zip(times_A, quals_A) if q >= tau]
    tB = [t for t, q in zip(times_B, quals_B) if q >= tau]
    if not tA or not tB:
        return math.inf
    return min(tA) / min(tB)
\end{lstlisting}

\subsection{Workflow audit utilities}
\begin{lstlisting}[language=Python, caption={Audit-and-bottleneck summarization with mean stage shares and drift statistics.}, label={lst:audit_fn}]
from typing import Sequence, Mapping, Optional, Dict, Any, List, Tuple
import numpy as np

def audit_bottlenecks(
    stage_durations: Sequence[Mapping[str, float]],
    top_k: int,
    drift_samples_ppm: Optional[Sequence[float]] = None
) -> Tuple[List[Tuple[str, float]], Optional[float], Optional[float]]:
    stages = sorted({s for run in stage_durations for s in run})
    shares_sum = {s: 0.0 for s in stages}
    for run in stage_durations:
        T = sum(run.get(s, 0.0) for s in stages)
        if T <= 0:
            continue
        for s in stages:
            shares_sum[s] += run.get(s, 0.0) / T
    R = max(1, len(stage_durations))
    mean_shares = {s: shares_sum[s] / R for s in stages}
    topk = sorted(mean_shares.items(), key=lambda kv: kv[1], reverse=True)[:top_k]
    mean_drift = p95_drift = None
    if drift_samples_ppm:
        arr = np.asarray(list(drift_samples_ppm), dtype=float)
        mean_drift = float(np.mean(arr))
        p95_drift = float(np.percentile(arr, 95))
    return topk, mean_drift, p95_drift
\end{lstlisting}

\subsection{Synthetic BQP demonstration (end-to-end)}
\begin{lstlisting}[language=Python, caption={Toy BQP demo (n = 24, 50 ms budget) comparing classical SA vs.~a greedy ``quantum-like'' heuristic.}, label={lst:toy_demo}]
from dataclasses import dataclass
from typing import Dict, Any, Tuple, Callable
import numpy as np, time, pandas as pd

@dataclass
class Budget:
    time_s: float
    cost_usd: float = 0.0
    energy_j: float = 0.0

def qubo_instance(n: int, density: float = 0.25, seed: int = 0):
    rng = np.random.default_rng(seed)
    Q = rng.normal(size=(n, n))
    Q = np.triu(Q, 1)
    Q = (Q + Q.T) * (rng.random(size=(n, n)) < density)
    return Q

def objective(Q: np.ndarray, x: np.ndarray) -> float:
    return float(x @ Q @ x)

def quality_from_objective(f: float, f_best: float, f_worst: float) -> float:
    if f_worst == f_best:
        return 1.0
    return float(np.clip((f_worst - f) / (f_worst - f_best), 0.0, 1.0))

def classical_sa_solver(inst, budget: Budget, seed: int):
    Q = inst
    rng = np.random.default_rng(seed)
    n = Q.shape[0]
    x = rng.integers(0, 2, size=n).astype(int)
    f = objective(Q, x)
    f_best = f_worst = f
    t_end = time.perf_counter() + budget.time_s
    while time.perf_counter() < t_end:
        i = int(rng.integers(n))
        x_try = x.copy()
        x_try[i] ^= 1
        f_try = objective(Q, x_try)
        if f_try <= f or rng.random() < 0.01:
            x, f = x_try, f_try
            f_best = min(f_best, f)
            f_worst = max(f_worst, f)
    q = quality_from_objective(
        f_best,
        f_best,
        (f_worst if f_worst > f_best else f_best + 1.0)
    )
    return q, {"time_s": budget.time_s, "energy_j": 0.0, "cost_usd": 0.0}

def quantum_like_greedy(inst, budget: Budget, seed: int):
    Q = inst
    rng = np.random.default_rng(seed)
    n = Q.shape[0]
    x = np.zeros(n, dtype=int)
    f = objective(Q, x)
    f_best = f_worst = f
    t_end = time.perf_counter() + budget.time_s
    while time.perf_counter() < t_end:
        gains = []
        for i in range(n):
            x_try = x.copy()
            x_try[i] ^= 1
            gains.append(objective(Q, x_try) - f)
        i_best = int(np.argmin(gains))
        if gains[i_best] < 0:
            x[i_best] ^= 1
            f = objective(Q, x)
            f_best = min(f_best, f)
            f_worst = max(f_worst, f)
        else:
            # reset to a new random bitstring if no improvement
            x = rng.integers(0, 2, size=n).astype(int)
            f = objective(Q, x)
    q = quality_from_objective(
        f_best,
        f_best,
        (f_worst if f_worst > f_best else f_best + 1.0)
    )
    return q, {"time_s": budget.time_s, "energy_j": 0.0, "cost_usd": 0.0}
\end{lstlisting}

\subsection{Bootstrap confidence intervals}
\begin{lstlisting}[language=Python, caption={Paired bootstrap CI for $S_{\text{norm}}(\tau)$}, label={lst:bootstrap_ci}]
import numpy as np

def speedup_ci_paired_bootstrap(
    df,
    tau: float,
    n_boot: int = 1000,
    alpha: float = 0.05
):
    rng = np.random.default_rng(42)
    inst_ids = sorted(df["instance_id"].unique())
    boot = []

    for _ in range(n_boot):
        sample = rng.choice(inst_ids, size=len(inst_ids), replace=True)
        sdf = df[df["instance_id"].isin(sample)]
        tA = sdf[sdf.solver == "Classical-SA"]["time_s"].to_list()
        qA = sdf[sdf.solver == "Classical-SA"]["quality"].to_list()
        tB = sdf[sdf.solver == "Quantum-like"]["time_s"].to_list()
        qB = sdf[sdf.solver == "Quantum-like"]["quality"].to_list()
        boot.append(normalized_speedup_at_tau(tA, qA, tB, qB, tau))

    lo, hi = np.quantile(boot, [alpha/2, 1 - alpha/2])
    return float(lo), float(hi)
\end{lstlisting}

\subsection{Timing utilities}
\begin{lstlisting}[language=Python, caption={Lightweight budget guard for loops; check guard.expired() inside kernels.}, label={lst:time_guard}]
import time

class TimeBudgetGuard:
    def __init__(self, seconds: float):
        self.deadline = time.perf_counter() + float(seconds)

    def expired(self) -> bool:
        return time.perf_counter() >= self.deadline

# Usage:
# guard = TimeBudgetGuard(budget.time_s)
# while not guard.expired():
#     ... do work ...
\end{lstlisting}

\begin{lstlisting}[language=Python, caption={Decorator to time named stages and emit a per-run dictionary for audit.}, label={lst:timed_stage}]
import time
import functools

def timed_stage(stage_name: str, sink: dict):
    def deco(fn):
        @functools.wraps(fn)
        def wrapper(*args, **kwargs):
            t0 = time.perf_counter()
            try:
                return fn(*args, **kwargs)
            finally:
                sink[stage_name] = sink.get(stage_name, 0.0) + (
                    time.perf_counter() - t0
                )
        return wrapper
    return deco
\end{lstlisting}

\end{document}